\documentclass[11pt]{article}

\usepackage[utf8]{inputenc}
\usepackage[T1]{fontenc}
\usepackage{amsmath,amssymb}
\usepackage{graphicx}
\usepackage{booktabs}
\usepackage[hypertexnames=false]{hyperref}
\usepackage{xurl}
\usepackage{xspace}
\usepackage{microtype}
\usepackage{xcolor}
\usepackage[margin=1in]{geometry}
\usepackage{float}

\graphicspath{{../figures/}{figures/}}

\hypersetup{
    colorlinks=true,
    linkcolor=blue,
    urlcolor=cyan,
    citecolor=blue,
    pdftitle={The Linux IOCTL Census: A Source-Derived Database of the Kernel Control-Code Surface},
    pdfauthor={Michael J. Bommarito II},
    pdfsubject={Vulnerability research, Linux kernel, ioctl, attack surface, static analysis},
    pdfkeywords={Linux kernel, ioctl, attack surface, static analysis, reachability, control-code surface, census}
}

\newcommand{\nmodules}{878\xspace}
\newcommand{\ndisp}{2{,}914\xspace}          
\newcommand{\niocdisp}{586\xspace}           
\newcommand{\niocmods}{337\xspace}           
\newcommand{\nsubtrees}{169\xspace}          
\newcommand{\nsubtreesioc}{104\xspace}       
\newcommand{\ncodes}{1{,}289\xspace}         
\newcommand{\ncmdsyms}{1{,}614\xspace}       
\newcommand{\cmdrate}{80\%\xspace}           
\newcommand{\nsinks}{3{,}583\xspace}         
\newcommand{\nunsanit}{3{,}201\xspace}
\newcommand{\ngates}{1{,}298\xspace}         
\newcommand{\nreach}{281\xspace}             
\newcommand{\ncapgated}{50\xspace}           
\newcommand{\nfiles}{338\xspace}             
\newcommand{\narmonly}{1\xspace}             
\newcommand{\nsrccov}{84\%\xspace}           
\newcommand{\nsurfcov}{99.7\%\xspace}     
\newcommand{\ncve}{22\xspace}
\newcommand{\ncvecov}{22\xspace}             
\newcommand{\nhandfound}{7\xspace}
\newcommand{\ncmdfound}{6\xspace}
\newcommand{\nctrl}{7\xspace}                
\newcommand{\nctrlfired}{4\xspace}           
\newcommand{\sys}{\textsc{Extract-Gate-Rank}\xspace}
\newcommand{\code}[1]{\texttt{#1}}

\title{\textbf{The Linux IOCTL Census}:\\
  A Source-Derived Database of the\\
  Linux Kernel Control-Code Surface}

\author{
Michael J.\ Bommarito II\thanks{Portions of this work were prepared with
assistance from large language models. The author is solely responsible for
all content, including any errors or omissions. This work was conducted for
defensive and authorized vulnerability-research purposes; see the data-release
and ethics notes in Section~\ref{sec:discussion}.}\\
\texttt{michael.bommarito@gmail.com}
}

\date{June 2026}

\begin{document}
\maketitle

\begin{abstract}
The \code{ioctl} system call is Linux's catch-all device-control interface. A
userspace program opens a device node and hands the driver a numeric command code
and an argument buffer, and the driver does whatever that code means, whether
configuring hardware, reading back state, or moving data into and out of the kernel.
Drivers define these commands themselves, by the thousand, and parse their arguments
in kernel context, which makes ioctl handlers one of the broadest and least uniform
local attack surfaces in the kernel. A handler that trusts an argument length it never
validates can read or write kernel memory out of bounds, and the command space is
catalogued in no central place. We present the Linux IOCTL Census, a source-derived and
queryable inventory of that surface. An \code{allmodconfig} build compiles \nmodules\
modules across \nsubtrees\ subtrees, and over them a single deterministic libclang
pass over the kernel source recovers \niocdisp\ ioctl dispatch entry
points, \ncodes\ decoded \code{\_IOC} command codes, \nsinks\ controlled-input sinks,
and \ngates\ permission gates. A second pass encodes the kernel's own documented
threat model as a queryable column, separating the capability-ungated ioctl surface,
an upper bound on unprivileged reach rather than proven reach, from the part a hard
capability gate puts out of scope. We backtest the
census against \ncve\ recent in-tree ioctl CVEs and release the structural tier as open
data, on a schema shared with the companion Windows IOCTL Census so a single query
spans both operating systems.
\end{abstract}

\section{Introduction}
\label{sec:intro}

A Linux \code{ioctl} call routes caller-supplied bytes to a driver handler that runs
in kernel context. A process opens a device node, hands the kernel a command number
and an argument pointer, and the kernel dispatches to the driver handler bound to
\code{file\_operations.unlocked\_ioctl} (or \code{compat\_ioctl}, or
\code{proc\_ops.proc\_ioctl}), which then runs over those bytes. A length
the handler trusts but never checks can become an out-of-bounds read or write. What makes the
surface hard to reason about is not any one handler but its shape in aggregate. A
single character device can carry dozens of commands, each with its own argument
layout and bounds assumptions, all written as \code{switch} arms behind a function
pointer and enumerated in no central place. The kernel even writes down which
crossings of this surface count as security bugs, in
\code{Documentation/process/threat-model.rst}, but that boundary is prose, not
something a tool can query.

Two existing approaches reach the surface from opposite ends. Dynamic scanners like
syzkaller and the syzbot infrastructure~\cite{syzkaller} drive handlers from written
and inferred descriptions and catch a large share of real kernel bugs, including the
race and runtime-state defects no static pass will reach. But a fuzzer sees only the
surface it is told about and happens to open, and it reports the bugs it triggered,
not the command space it never touched. Curation takes the other side. The
unprivileged-reach LPE catalogue and the threat-model documentation record which entry
points the community treats as in scope, but they are hand-kept lists, not a
systematic enumeration of the decoded command space. What is missing is the surface
itself, persisted and queryable.

That gap is what the census fills, and it fills it from source. A libclang knowledge
graph of the kernel tree lets us read the static command space directly, the
dispatchers, the decoded commands, the handlers, the controlled-input sinks, and the
permission gates, as one row per \code{(module\_sha256, symbol)}, for \nmodules\
modules. We make three contributions.

\begin{enumerate}
  \item \textbf{A source-derived ioctl census over the compiled tree.} A queryable inventory
    of dispatchers, decoded command codes, handlers, controlled-input sinks, and
    permission gates over every handler-bearing module an \code{allmodconfig} build
    compiles (\nmodules\ modules, \nsubtrees\ subtrees). Rebuilding for a second
    architecture (\code{arm64}) adds only \narmonly\ ioctl module and no new command
    codes (Section~\ref{sec:eval}).
  \item \textbf{A threat-model-as-data reachability filter.} A source-side
    necessary-condition filter that encodes the kernel's documented threat model as a
    queryable column, turning ``does a hard capability gate block this path'' into
    a query (Section~\ref{sec:method}). It is an upper-bound filter, not a
    reachability classifier.
  \item \textbf{A tiered, validated dataset.} We release a public structural tier
    under an open license and withhold the targeting tier
    (Section~\ref{sec:discussion}). It uses the same schema as our companion Windows
    IOCTL Census, so one query spans both, and it ships with a CVE-recall pass whose
    every miss is attributed to a stated design bound (Section~\ref{sec:backtest}).
\end{enumerate}

This is a measurement and dataset paper, not a bug-finding paper. Three
limitations matter, and we state them up front rather than bury them. The reachability
filter is a necessary condition on the capability axis, never proven
node-openability, so every ``user-reachable'' count is an upper bound. The
sanitized signal is a heuristic, a bounds check earlier in the same source function,
not a proof that the check guards the sink on every path. And
coverage is what an \code{x86\_64} \code{allmodconfig} build compiles (\nsrccov\ of
in-tree source files; the uncompiled remainder is dominated by other-architecture
platform code, Section~\ref{sec:eval}). Every headline number is read from
the built census and the validation artifacts; a number that does not survive the
validity pass (Section~\ref{sec:backtest}) is removed, not softened.

\section{Method}
\label{sec:method}

\begin{figure}[t]
  \centering
  \includegraphics[width=\textwidth]{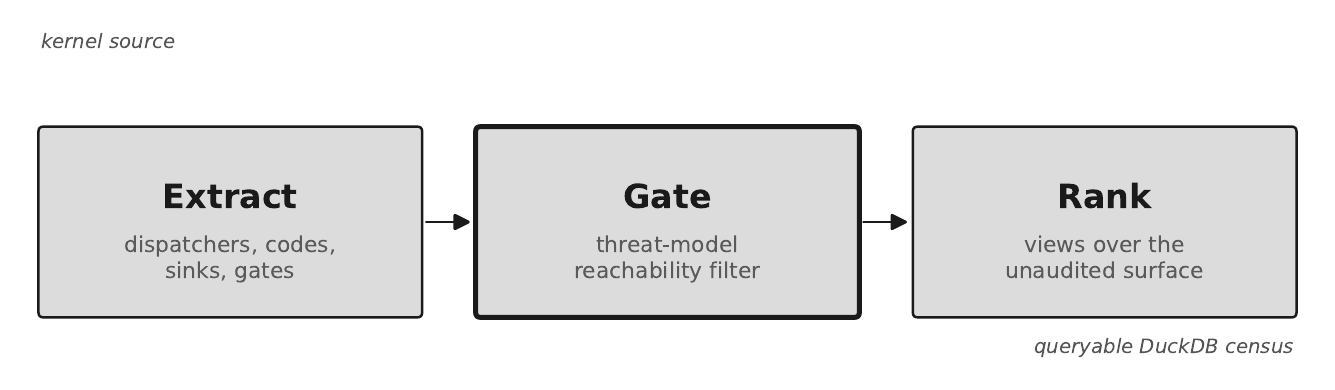}
  \caption{The \sys\ pipeline. \emph{Extract} recovers dispatch, decoded
  \code{\_IOC} codes, controlled-input sinks, and permission gates from kernel
  source with no fuzzing or symbolic execution. \emph{Gate} applies the
  threat-model reachability filter (a necessary-condition upper bound, not a
  classifier). \emph{Rank} serves views over the unaudited reachable surface; the
  LLM enrichment layer is optional and off the critical path.}
  \label{fig:pipeline}
\end{figure}

\sys\ reads the kernel source and materializes a single relational store. We organize
the work into three stages of our own naming (Figure~\ref{fig:pipeline}), sharing the
key \code{(module\_sha256, symbol)}. \emph{Extract}, the deterministic core, recovers
each module's dispatch surface, decoded commands, controlled-input sinks, and
permission gates from a libclang-derived graph of the source. \emph{Gate} turns the
kernel's documented reachability boundary into a queryable column. \emph{Rank} builds
the views that surface the unaudited modules first.

 A \emph{module} throughout this paper is one kernel source file that
registers a handler in an ops struct, identified by
\code{module\_sha256 = md5(arch : tree-relative path)} rather than by any compiled
\code{.ko}; the same file built for two architectures is therefore two modules. The
key column is named \code{sha256} for cross-OS schema compatibility, where the
companion Windows census stores a binary content hash; the Linux census, with no
compiled artifact to hash, stores the path MD5 in that column. The
store is keyed by \code{(module\_sha256, symbol)}: a stable 64-bit
\code{hash(USR)} addresses each row, and companion \code{name}/\code{file}/\code{line}
columns carry the human location with a \code{loc\_kind} of \code{symbol} or
\code{site}. Because Linux source is relocatable, \code{loc\_kind} takes the place of
the load address a binary census would store; this is what lets the schema match our
companion Windows census's \code{(binary\_sha256, function\_va)} keying.
Table~\ref{tab:schema} lays out the per-stage tables, the key they share, and the
views built over them.

\subsection{Extract}
\label{sec:extract}
\begin{figure}[t]
  \centering
  \includegraphics[width=\linewidth]{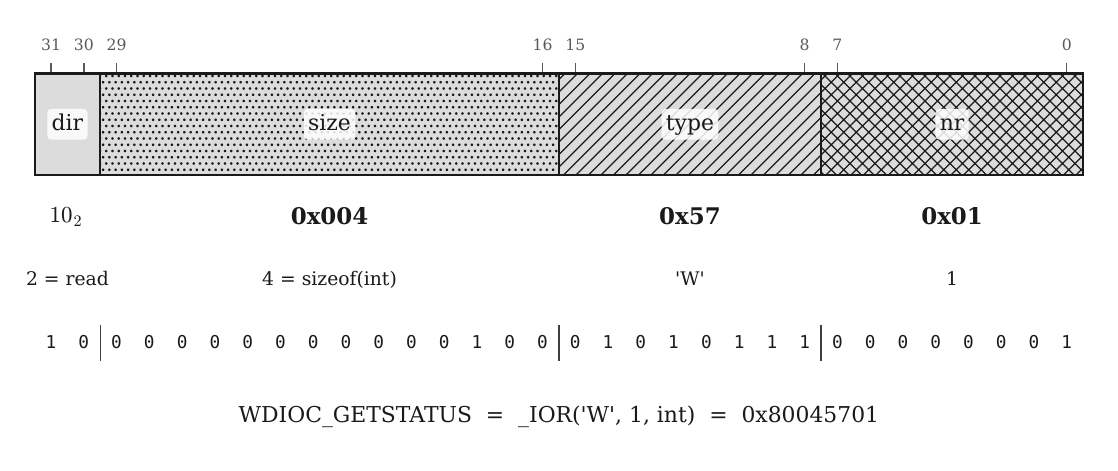}
  \caption{The resolver recovers the asm-generic \code{\_IOC} command-code encoding, a
  32-bit number split from the top bit down into \code{dir} (direction, bits 31:30),
  \code{size} (argument size, 29:16), \code{type} (magic byte, 15:8), and \code{nr}
  (command number, 7:0). The worked example is the pinned resolver anchor
  \code{WDIOC\_GETSTATUS = \_IOR('W', 1, int) = 0x80045701}, whose read direction
  transfers data from the kernel to the user buffer.}
  \label{fig:ioc_layout}
\end{figure}

The deterministic pass is built over what we call a libclang-derived knowledge graph
of the source, the relational index libclang produces from each translation unit. It
holds the functions and their ops-struct bindings, the call edges between them, and
the decoded command codes, sinks, and gates attached to each handler, all addressed by
a USR-based symbol table. The call graph is one layer of this index, not the whole of
it. The index is extracted per subtree from an \code{allmodconfig}
\code{compile\_commands.json}, sharded for parallelism but merged so the analysis sees
the whole compiled tree. The present
build covers every module the \code{x86\_64} \code{allmodconfig} configuration
compiles (\nmodules\ modules across \nsubtrees\ subtrees). The substrate is pinned at a single
\code{git describe} value (\code{v7.0-rc7}) recorded in the \code{kb\_meta} table,
alongside the clang version and the build configuration. A rebuild against a later
kernel then records the revision it actually saw, not a value typed into prose. For each module the pass records, in order:

\paragraph{Dispatch entries and decoded codes.} We identify ioctl dispatchers as the
functions bound to the ioctl fields of an ops struct (\code{unlocked\_ioctl},
\code{compat\_ioctl}, or \code{proc\_ioctl}), then walk the call
closure to the per-command handler bodies. The genuinely new component decodes the
command codes. A \code{\_IOC} resolver expands the command macros via
\code{clang -E -dM} against the uapi headers seen by the same compile database. It
then decodes each \code{switch}-case label appearing in a dispatcher or handler into
the four fields of the kernel's \code{\_IOC} encoding~\cite{ioc_uapi}, its direction
(none/read/write/read-write), magic byte, ordinal, and argument-struct size
(Figure~\ref{fig:ioc_layout}). Each decoded code is linked
to its handler address.

\paragraph{Sinks and gates.} For each handler we record two things. The
controlled-input sinks are the user-to-kernel copy primitives
(\code{copy\_from\_user}, \code{\_\_copy\_from\_user}, \code{get\_user},
\code{\_\_get\_user}, \code{memdup\_user}, \code{strncpy\_from\_user},
\code{copy\_from\_iter}, and \code{import\_ubuf}) reachable within a fixed call-hop
window, a depth-six closure over the resolved call graph. The permission gates on the
path are capability checks (\code{capable} and \code{ns\_capable}, classified as
init-namespace or any-user-namespace), LSM hooks, \code{f\_mode} and field checks, and
fd lookups. A sink is marked
\emph{unsanitized} precisely as ``no preceding hard bounds field-check in the same
function.'' That is a textual-order heuristic, a check earlier in the source, not a
control-flow proof that the check guards every path to the sink. It can over-count a
sink as sanitized when the preceding check is on a different value or a non-taken
branch, and it does not model inter-procedural guards. We report the proxy as a structural counter and treat the
unsanitized total as an upper bound on attention, not a measure of guardedness or a
defect count. Its precision is not yet manually audited.

The Extract pass is deterministic and reproducible. The structural tables carry no
ordering or wall-clock nondeterminism and rebuild byte-identically from the pinned
input (Section~\ref{sec:eval}). It is bounded, not exhaustive. The
dispatcher-recovery pass sees handlers bound to \code{file\_operations} and
\code{proc\_ops}; tty line disciplines, core-switch helpers, and netlink-style
command interfaces are out of the current registration model
(Section~\ref{sec:backtest}).

\subsection{Gate}
\label{sec:gate}
\begin{figure}[t]
  \centering
  \includegraphics[width=0.86\linewidth]{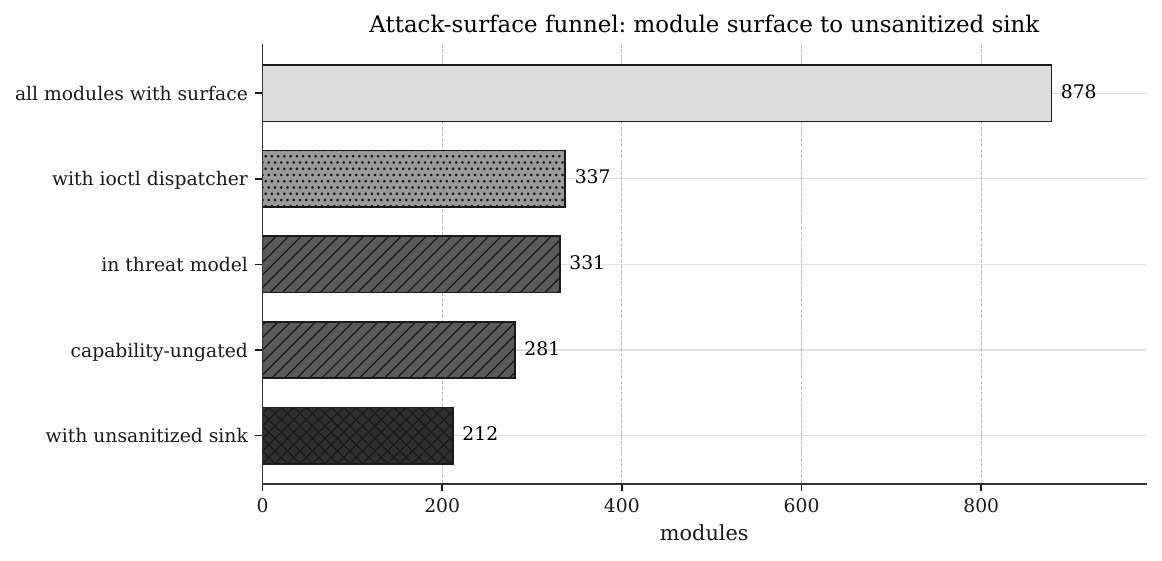}
  \caption{The filter, not a classifier, does the narrowing. Of the \niocmods\
  ioctl modules the threat-model filter excludes \ncapgated\ behind a hard
  capability gate and admits \nreach\ as the capability-ungated upper bound on
  unprivileged reach, of which a majority carry at least one unsanitized
  controlled-input sink under the line-precedence proxy. Both stages are upper
  bounds, not reach or defect counts.}
  \label{fig:funnel}
\end{figure}

The reachability column is a necessary-condition upper-bound filter, not a
classifier. A module is admitted (\code{capability\_ungated}) when its ioctl path
carries no hard init-namespace capability gate, and \code{in\_threat\_model} unless
it is a staging driver or otherwise carved out by the kernel's documented model.
The carve-outs come straight from the kernel's own threat-model and security-bugs
process documentation~\cite{linuxthreatmodel}: init-namespace
\code{CAP\_SYS\_ADMIN}-gated paths, debugfs, and staging drivers are out of model; a
bug reachable only that way is generally not treated as a kernel security
vulnerability. The filter discriminates on the capability axis. Of the \niocmods\
modules that carry an ioctl dispatcher, it excludes \ncapgated\ behind a hard
init-namespace capability gate and admits \nreach\ as the capability-ungated upper
bound on unprivileged reach. The remaining 6 are staging drivers, out of model; the
admitted surface spans \nsubtreesioc\ subtrees.
Figure~\ref{fig:funnel}
shows the funnel.

What the filter cannot do is detect node-DAC-only restrictions, a device node that
is root- or hardware-only by virtue of its udev/devtmpfs mode but carries no
in-handler \code{capable()} check. We therefore never equate an admitted module
with an unprivileged-reachable one. A \code{capability\_ungated} verdict means ``no
hard capability gate blocks this path,'' and \nreach\ is an explicit upper bound on
unprivileged reach. The node-DAC extractor (registration call plus udev/devtmpfs
default modes) is incomplete, so reach that depends on node DAC is not resolved;
where the node mode is unknown the census says so rather than guessing. The
predicted node-DAC false-positive class and its measured cost are stated in
Section~\ref{sec:backtest}. Every ``user-reachable'' count in this paper is this
capability-ungated upper bound, and we say so each time.

\subsection{Rank}
\label{sec:rank}
Rank is a set of relational views over the store. One sorts the
capability-ungated dispatchers carrying unsanitized controlled-input sinks by how
many such sinks each has, dropping any a filed finding already covers. This
ranking is an unvalidated prioritization heuristic. It rests on the unsanitized-sink
proxy, whose precision we do not audit (Section~\ref{sec:backtest}), so it orders
attention rather than asserting defects. A
coverage view rolls up, per subtree, how much surface was recovered and how much
the filter admits. A cross-module view groups the same decoded
\code{(type, ordinal, access)} triple across distinct modules, surfacing shared
control interfaces such as the generic VFS ioctls (Section~\ref{sec:eval}). Every
view is a read-only query, so it can run against a store while a build is still
populating it.

\paragraph{Optional enrichment, off the census claims.} The pipeline includes an
exploratory, source-grounded LLM triage layer that classifies a
reachability-filtered subset of handlers under a safe-refute default and a
citation-grounding validator that rejects any verdict not anchored to a cited line.
It is the source-side counterpart of our companion work on LLM-assisted target
selection for stripped Windows binaries~\cite{needles}, applied here to kernel source.
On the current-tree prioritizer it returned all-low, which alone says little. There, an
all-low result from this layer is consistent both with a clean reachable subset and
with context too thin to surface anything. To calibrate it we ran a positive control:
for the backtest CVEs the census found, we fed the layer the \emph{pre-fix}
source of the handler from git. It fired medium-or-higher on \nctrlfired\
of \nctrl\ vulnerable revisions, and on the high-confidence cases it named the
correct primitive (an out-of-bounds write for a board-index ioctl, out-of-bounds
reads for two BAR-index ioctls). The \nctrl\ minus \nctrlfired\ misses are
dispatcher-level. The vulnerability lives in a callee below the handler body the
layer was shown. We report this as a calibration signal on the optional layer, not a
detector claim; the structural census needs no model, and the enrichment outputs stay
in the withheld targeting tier (Section~\ref{sec:discussion}).

\begin{table}[t]
  \centering
  \small
  \begin{tabular}{@{}ll@{}}
    \toprule
    \multicolumn{2}{@{}l}{\textbf{Store schema}\quad key \code{(module\_sha256, symbol)}} \\
    \midrule
    \emph{Extract} & \code{modules}, \code{dispatchers}, \code{ioctl\_codes}, \code{handlers}, \\
                   & \code{functions}, \code{call\_edges}, \code{gates}, \\
                   & \code{taint\_sinks}\textsuperscript{\dag} \\
    \emph{Gate}    & \code{reachability} (\code{in\_threat\_model}, \code{capability\_ungated}) \\
    \emph{Enrich}  & \code{handler\_enrichment}\textsuperscript{\dag} \\
    \midrule
    \multicolumn{2}{@{}l}{\textbf{Views}} \\
    \code{v\_unaudited\_unpriv\_surface}\textsuperscript{\dag} & ranked unaudited surface \\
    \code{v\_coverage}                  & recovered/admitted per subtree \\
    \code{v\_cross\_build}              & shared \code{(type, ordinal, access)} across modules \\
    \bottomrule
  \end{tabular}
  \caption{The relational store, keyed throughout by \code{(module\_sha256, symbol)},
  the source-side analogue of the Windows census's \code{(binary\_sha256,
  function\_va)}, with a \code{loc\_kind} of \code{symbol} or \code{site} standing in
  for the load address. The \textsuperscript{\dag} rows form the withheld targeting
  tier; the rest ship in the public structural dataset.}
  \label{tab:schema}
\end{table}

\section{Anatomy of the Recovered Surface}
\label{sec:eval}

We read the census back rather than score it against a detection target: what the
deterministic pass recovered, the fraction the threat-model filter admits, and the
points where recovery stops. Table~\ref{tab:corpus} gives the totals over the
\code{x86\_64} \code{allmodconfig} tree.

\begin{table}[t]
  \centering
  \small
  \setlength{\tabcolsep}{6pt}
  \begin{tabular}{@{}l r r@{}}
    \toprule
                                          & \textbf{Count} & \textbf{} \\
    \midrule
    \multicolumn{3}{@{}l}{\textbf{Coverage} of \nsubtrees\ in-tree subtrees} \\
    Subtrees with an ioctl handler        & \nsubtreesioc  &      \\
    Modules registering a handler         & \nmodules      &      \\
    Modules with an ioctl dispatcher      & \niocmods      &      \\
    \addlinespace
    \multicolumn{3}{@{}l}{\textbf{Recovered surface}} \\
    Registered dispatcher entries         & \ndisp         &      \\
    \quad \code{unlocked}/\code{compat\_ioctl} dispatchers & \niocdisp &  \\
    Decoded \code{\_IOC} commands         & \ncodes        & of \ncmdsyms \\
    Controlled-input taint sinks          & \nsinks        &      \\
    \quad unsanitized (line-precedence proxy) & \nunsanit  &      \\
    Permission gates                      & \ngates        &      \\
    \addlinespace
    \multicolumn{3}{@{}l}{\textbf{Reachability filter} (per ioctl module)} \\
    Hard-capability-gated (excluded)      & \ncapgated     &      \\
    Capability-ungated (upper bound)      & \nreach        &      \\
    \bottomrule
  \end{tabular}
  \caption{The corpus. Counts are over the full \code{x86\_64} \code{allmodconfig}
  tree. Command resolution is \cmdrate\ (\ncodes\ of \ncmdsyms\ \code{switch}-case
  symbols decoded); the unresolved tail is dominated by legacy non-\code{\_IO*}
  numeric command constants that carry no decodable structure. The unsanitized count is a proxy counter (Section~\ref{sec:method}),
  and \nreach\ is a capability-ungated upper bound on unprivileged reach, not a
  reach count.}
  \label{tab:corpus}
\end{table}

\paragraph{What the deterministic pass alone yields.} Across the \nsubtrees\
subtrees the \emph{Extract} stage recovers \nmodules\ modules that register a
handler and \ndisp\ registered dispatcher entries, of which \niocdisp\ are ioctl dispatchers across \niocmods\ modules. It decodes \ncodes\ \code{\_IOC} commands, extracts \nsinks\
controlled-input sinks, and records \ngates\ permission gates. Because all of it is
stored together, the census answers structural questions a bug list cannot. The
cross-module view groups one decoded \code{(type, ordinal, access)} triple
across distinct modules, and the reachability column tells a handler behind a hard
capability gate from one with none. None of this needs the model.

\paragraph{Coverage and a second architecture.} The build compiles \nsrccov\ of the
in-tree \code{.c} files; the uncompiled remainder is dominated by other-architecture
platform code and by drivers \code{allmodconfig} cannot co-enable. To see how much a
single architecture leaves out, we built the census a second time for \code{arm64} and
unioned the two on the shared key. Across \code{x86\_64} and \code{arm64} the census
spans \nfiles\ distinct ioctl-bearing modules; exactly \narmonly\ is \code{arm64}-only,
the Apple \code{adp\_drv} DRM driver, and the \code{arm64} build adds no new distinct
\code{(type, ordinal)} command codes. The \code{x86\_64} build alone thus accounts for
\nsurfcov\ of the ioctl-bearing modules the two architectures recover between them.

\paragraph{The surface is character-device-dominated.} The analyzed modules are
overwhelmingly character devices, with a small block-device, sysfs, and proc tail.
The dominance is expected, since \code{unlocked\_ioctl} on a misc or character node
is the canonical local ioctl entry point, but it also tells an auditor where the
surface lives.

\paragraph{Command resolution and the decoded space.} The \code{\_IOC} resolver
decodes \ncodes\ of \ncmdsyms\ \code{switch}-case command symbols, a resolution
rate of \cmdrate. The unresolved tail is
dominated by legacy non-\code{\_IO*} numeric command constants that carry no
decodable structure, with a small residual class of case labels in files still
absent from the \code{allmodconfig} compile database; we report both as a category
rather than dropping them. Hand-verified constants, including \code{WDIOC\_GETSTATUS = 0x80045701}, are pinned
in unit tests as resolver anchors and a compile oracle checks the resolved set
against the uapi headers.

\begin{figure}[tbp]
  \centering
  \includegraphics[width=0.92\linewidth]{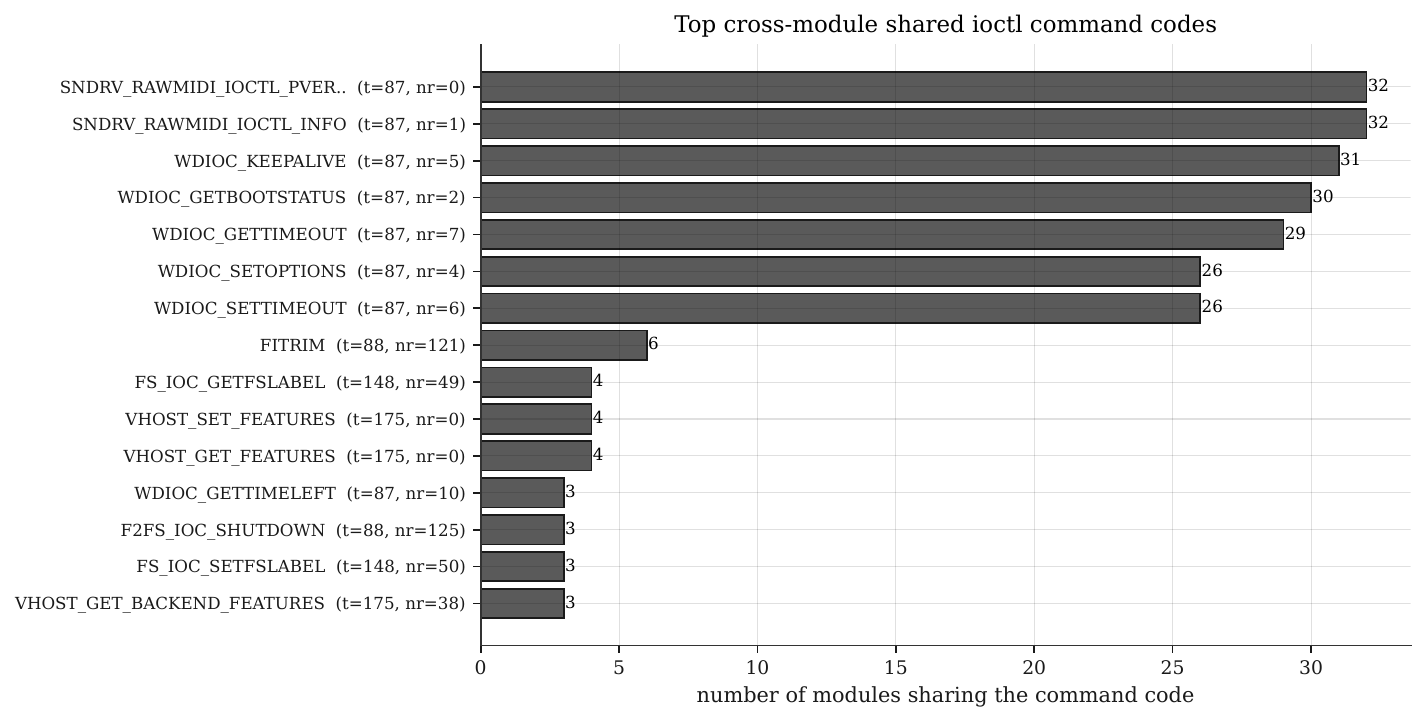}
  \caption{Top cross-module shared decoded \code{(type, ordinal, access)} triples: the
  watchdog \code{'W'}-type commands recur across the 31 watchdog drivers (32 counting
  an ALSA rawmidi driver that collides on the same magic byte), from the
  \code{v\_cross\_build} view.}
  \label{fig:cross_module_cmds}
\end{figure}

\paragraph{Shared commands recur across modules.}
Figure~\ref{fig:cross_module_cmds} ranks the \code{(direction, type, ordinal)}
encoding triples that recur across modules. We group by this triple rather than the
full 32-bit \code{\_IOC} value, ignoring the argument-size field, which surfaces both
driver-class command reuse and magic-byte collisions. The top recurrences are the
watchdog control interface: the
\code{WDIOC\_*} commands, all of magic type \code{'W'}, recur across the 31 watchdog drivers,
with \code{WDIOC\_GETSUPPORT}, \code{WDIOC\_GETSTATUS}, and \code{WDIOC\_KEEPALIVE}
each appearing in 31 modules. Grouping by the decoded triple rather than the symbol
pushes the top counts to 32 by surfacing a magic-byte collision. The \code{'W'} type
with ordinal 1 is used by both \code{WDIOC\_GETSTATUS} (watchdog) and
\code{SNDRV\_RAWMIDI\_IOCTL\_INFO} (ALSA rawmidi). They share that triple but encode
different argument sizes, so they are in fact different 32-bit codes
(\code{0x80045701} versus \code{0x80005701}) that collide only on the magic byte and
ordinal. These are shared driver-class contracts, not copy-paste accidents. Still, a bug in a
shared command pattern tends to recur across the drivers that implement it, so the
cross-module view prioritizes a single audit that covers several modules at once.

\begin{figure}[tbp]
  \centering
  \includegraphics[width=0.86\linewidth]{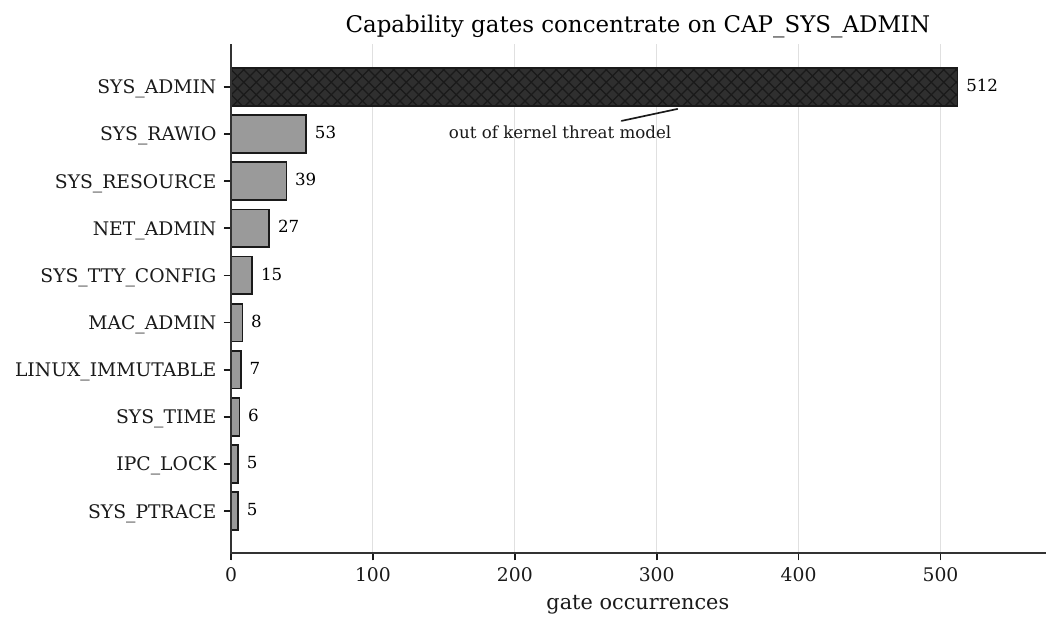}
  \caption{Permission gates by capability. \code{CAP\_SYS\_ADMIN} dominates by an
  order of magnitude, and it is exactly the capability the kernel threat model treats
  as out of model, so much of the gated surface sits outside the in-model attack
  surface.}
  \label{fig:cap_gates}
\end{figure}

\paragraph{Gates concentrate on the one over-broad capability.} The \ngates\
permission gates are dominated on the capability axis by \code{CAP\_SYS\_ADMIN}
(512 occurrences), with \code{CAP\_SYS\_RAWIO} (53) and \code{CAP\_SYS\_RESOURCE}
(39) distant second and third (Figure~\ref{fig:cap_gates}). The long \code{CAP\_SYS\_ADMIN} tail is itself
worth stating plainly. A large fraction of the gated ioctl surface is gated by the
single most over-broad capability, which is exactly the one the kernel threat model
treats as out of model. That places much of the gated surface outside the in-model
attack surface, and it is why the threat-model gate, not raw gate counts, is the
column that matters for reachability.

\paragraph{The analysis tables are deterministic.} The analysis tables are built
from the pinned input and addressed by \code{hash(USR)}, which carries no ordering
or wall-clock nondeterminism, so they rebuild deterministically. We verify this with
a twice-rebuild content-hash diff. Two independent rebuilds produced
\emph{byte-identical} tables (\code{modules}, \code{dispatchers},
\code{ioctl\_codes}, \code{handlers}, \code{gates}, and the withheld
\code{taint\_sinks}; provenance timestamps excluded). The public structural export
is the subset of these tables minus \code{taint\_sinks} and the other targeting-tier
rows (Table~\ref{tab:schema}). The check first caught two real nondeterminism
sources (a non-deterministic group aggregate and a primary-key collapse on handlers
registered for multiple file-operations fields), which we fixed; it is the release
gate and now passes. We record the build provenance in \code{kb\_meta} so a third
party can reproduce the structural counts from the pinned source.

\section{Backtest: Recovering Known CVEs}
\label{sec:backtest}

A source-side map is only as trustworthy as its coverage of the bugs already known.
We assembled \ncve\ recent in-tree Linux ioctl CVEs, each cross-checked against the
kernel \code{vulns.git} record and reduced to a known vulnerable handler, command
code, and subtree. For each, we asked what the deterministic pass recovered and
whether the threat-model filter admitted it. The point is a map of where the method
sees and where it is blind, not a recall score. Every miss is traced to a stated
design bound, and any number that does not survive the pass is dropped rather than
softened.

\paragraph{Coverage is no longer the bound.} All \ncve\ CVEs fall in census-covered
subtrees, so the \code{allmodconfig} build leaves no coverage gap, where an earlier
partial snapshot left several \code{drm}, \code{media}, and \code{fs} CVEs in
un-extracted paths. Closing the coverage gap is what lets the backtest cleanly isolate the one
remaining bound, the registration model. Table~\ref{tab:backtest} gives the per-stage
funnel over all \ncve\ CVEs.

\begin{table}[t]
  \centering
  \small
  \begin{tabular}{lrr}
    \toprule
    \textbf{Stage} & \textbf{Count} & \textbf{Denominator} \\
    \midrule
    In a covered subtree            & \ncvecov & \ncve \\
    Handler recovered               & \nhandfound & \ncvecov \\
    \quad command decoded           & \ncmdfound & \nhandfound \\
    \quad filter admits (TP)        & 5 & \nhandfound \\
    \quad filter admits (node-DAC FP) & 2 & \nhandfound \\
    \bottomrule
  \end{tabular}
  \caption{CVE backtest funnel. \textbf{In a covered subtree} is now all \ncve\ CVEs:
  the \code{allmodconfig} build leaves no coverage gap. \textbf{Handler recovered} excludes 15
  misses, every one a registration- or dispatch-model limitation: eight DRM/V4L2
  ops-table handlers, three tty line disciplines, and four functions reached only as
  callees below the dispatcher switch (\code{blk\_ioctl\_discard}, and the
  \code{cifs}/\code{nilfs2} sub-handlers). The filter admits all \nhandfound; the two
  node-DAC false positives are both \code{pci\_endpoint\_test}, the predicted class.
  Its capability-axis discrimination (the \ncapgated\ excluded on the full census) is
  not exercised by this subset.}
  \label{tab:backtest}
\end{table}

\paragraph{Every miss maps to a dispatch pattern the method does not model.} Of the
\ncve\ CVEs the dispatcher-recovery pass found the vulnerable handler for \nhandfound;
the 15 misses lead the result because, with coverage complete, every one is a
registration- or dispatch-model limitation rather than a missing subtree. The DRM and
V4L2 cores register per-command handlers through dispatch tables
(\code{drm\_ioctl\_desc}, \code{v4l2\_ioctl\_ops}): the core copies the argument and
passes a kernel pointer, so the per-command handler is neither a
\code{file\_operations} entry nor a \code{switch(cmd)} case the extractor reads. Eight
\code{drm}/\code{media} CVEs miss here. Three more are the tty line disciplines
(\code{vt\_ioctl}, \code{n\_gsm}, the TIOCLINUX selection path) registered through
\code{tty\_ldisc\_ops}. The last four are functions reached only as callees below the
registered dispatcher's \code{switch} rather than as the dispatcher itself, so the
dispatcher is recovered but the vulnerable callee is not. These are the block-core
helper \code{blk\_ioctl\_discard} (\code{BLKDISCARD}, \code{BLKSECDISCARD}) and the
\code{cifs}/\code{nilfs2} sub-handlers \code{cifs\_dump\_full\_key} and
\code{nilfs\_ioctl\_wrap\_copy}. The ops-table pattern (DRM, V4L2, ALSA) is the
largest remaining bound and the clear next target for the registration model.

\paragraph{Command resolution on the found set.} Conditional on the handler being
found, the \code{\_IOC} resolver decoded the command for \ncmdfound\ of \nhandfound.
The single miss is the applicom \code{cmd==6} read-board case, a raw numeric constant
with no \code{\_IO*} structure to decode. The \code{pci\_endpoint\_test}
\code{PCITEST\_*} and \code{fastrpc} codes decode correctly. \code{PCITEST\_BAR} is
\code{\_IO('P', 1)} and the \code{fastrpc} \code{\_IOCTL\_*} codes use the
\code{\_IO*} encoding.

\paragraph{What the reachability filter admits.} The filter is a necessary condition,
not a classifier, so we report what it admits and rejects rather than one accuracy
number. It admits all \nhandfound\ handler-found CVEs, finding no hard capability gate
on any of their paths. Two of those seven are deliberate node-DAC controls, both
\code{pci\_endpoint\_test}. Their device node is exposed only on CI and test hardware,
so they are effectively root-only, yet no \code{capable()} check gates the handler, so
the filter admits them. These are the false admits the method predicts, because the
filter reads the capability gates in the handler, not who is allowed to open the node.
On this subset, then, the filter makes five correct admits and two node-DAC false
admits. Its real discrimination is on the capability axis, the \ncapgated\ modules the
full census excludes for a hard gate; seven CVEs are too few and too uniform to test
the node-DAC axis at all. The two capability-gated CVEs in the corpus, \code{n\_gsm}
and \code{cifs}, fall in the handler-miss set and never reach the filter. The node-DAC
miss appears exactly where the method says it would, which is why we treat every
``user-reachable'' count as a capability-ungated upper bound.

\paragraph{What the backtest does not claim.} We report no precision audit of the
unsanitized-sink proxy. The backtest exercises handler recovery, command resolution,
and the capability-axis filter against known-vulnerable handlers; it does not
validate the sanitized/unsanitized split, for which we have no CVE-derived ground
truth and no manual precision audit. The unsanitized count stays a structural counter
and an upper bound on attention (Section~\ref{sec:method}). The optional enrichment
layer is reported separately as a calibration signal, not a census claim. A positive
control that fed it the pre-fix vulnerable source of the found CVEs fired
medium-or-higher on \nctrlfired\ of \nctrl\ revisions (Section~\ref{sec:method}).
That control pre-selects known-vulnerable handlers, so it measures best-case
sensitivity, not end-to-end detection.

\section{Discussion}
\label{sec:discussion}

\paragraph{Where this sits.} The recovery techniques are standard; the contribution
is that the surface persists as a queryable store rather than dissolving into a bug
list. syzkaller and syzbot~\cite{syzkaller} are the dominant dynamic approach to
kernel-surface testing, and they reach a class of bug a static pass cannot:
runtime-state, race, and reachable-only-at-execution paths. But a fuzzer reports the
bugs it triggers, not the shape of the command space it did not, and a command with
no description or behind an unopened node is never exercised. The census is the
complement: it enumerates and gates the entire decoded command space statically,
including commands no description covers, and leaves dynamic triggering to the
fuzzer. DIFUZE~\cite{difuze} makes that triggering interface-aware, recovering the
ioctl argument structure to drive a Linux driver fuzzer from the right input shape.
The census supplies the same command-and-argument inventory statically across the
whole tree, whether or not a harness for a given node exists. On the static side,
DR.~CHECKER~\cite{drchecker} runs soundy flow- and context-sensitive taint over
Linux driver entry points and Static Driver Verifier~\cite{sdv} model-checks driver
API contracts. Both analyze a driver deeply; the census instead enumerates and gates
which drivers and commands are reachable at all, the target list a deep tool needs.
General-purpose kernel static tools such as sparse, Smatch, Coccinelle, and
CodeQL-style source queries each enforce a bug pattern or semantic rule a check at a
time; the census is orthogonal, a persistent inventory of the dispatch surface such
checkers run against, not a checker itself. On the symbolic side, IOCTLance~\cite{ioctlance} vets individual Windows
WDM/WDF driver ioctl handlers for vulnerability patterns and POPKORN~\cite{popkorn}
automates the discovery of vulnerable Windows kernel drivers at scale with symbolic
execution. Both deepen the analysis of a handler where the census broadens the
inventory of all of them and tells a deeper tool where to look. Surveys of signed Windows drivers exposing privileged primitives
through ioctl~\cite{screweddrivers} characterize the bring-your-own-vulnerable-driver
surface and motivate the census-as-dataset framing. Our companion Windows IOCTL
Census~\cite{ioctlcensuswin} is the schema sibling of this work, our own artifact
rather than an external peer-reviewed source. Sharing a schema means the same query
runs over both, but the focus of this paper is the Linux surface itself. We combine
rather than invent; to our knowledge this is the first queryable, source-derived
ioctl census on this schema with the threat-model filter, not the first static look
at ioctl-like surfaces.

\paragraph{What a persistent store enables.} The value is that the structure persists.
Instead of re-scanning, an auditor gets a ranked, deduplicated worklist of the
capability-ungated, unsanitized-sink surface. The cross-module view groups one
decoded command into the cohort of drivers that share it, so a bug in a shared
command pattern is audited once across all of them. Encoding the kernel's threat
model as \code{in\_threat\_model} and \code{out\_of\_model\_reason} columns turns
``does any hard capability gate block this path'' from prose a reviewer remembers into
a query. The \code{CAP\_SYS\_ADMIN} dominance of Section~\ref{sec:eval} is one clean
consequence.

\paragraph{Limitations.} Each is a measured bound, several surfaced by the backtest.
Coverage is what an \code{x86\_64} \code{allmodconfig} build compiles, \nsrccov\ of
in-tree source files; the uncompiled remainder is dominated by other-architecture
platform code and by drivers \code{allmodconfig} cannot co-enable. The \code{arm64}
rebuild (Section~\ref{sec:eval}) adds \narmonly\ ioctl module and no new command
codes, which bounds how much a single configuration leaves out. Registration-type bounds are now the binding
limitation on what the dispatcher pass sees. Three patterns are out of the current
registration model: ops-table dispatch (the DRM, V4L2, and ALSA cores route
per-command handlers through tables such as \code{drm\_ioctl\_desc} and
\code{v4l2\_ioctl\_ops}, copying the argument in the core), tty line disciplines, and
below-dispatcher callee helpers. These are the entire handler-miss set in the backtest
now that coverage is complete, and the ops-table pattern is the largest single class. The reachability
filter is a necessary condition on the capability axis only, not proven
node-openability, and the two \code{pci\_endpoint\_test} false positives are exactly
the predicted node-DAC class. The sanitized signal is the same heuristic, a bounds
check earlier in the source rather than a proof it guards the sink, and its precision
is not yet manually audited.
The \nunsanit\ unsanitized count is a structural counter and an upper bound on
attention, not a defect count. The enrichment layer is an optional triage signal, not
part of the census claims. It returned all-low on the current build, while a positive
control on \nctrl\ pre-fix vulnerable handler revisions fired on \nctrlfired\
(Section~\ref{sec:method}). We keep it off the headline and in the withheld targeting
tier.
The census is also static. A \code{capability\_ungated} verdict is a property of the
call graph and the gate set, not a dynamic proof that an unprivileged process can open
the node and drive the handler. The census is the complement to a fuzzer, not a
replacement.

\paragraph{Ethics and data release.} Everything the census reads is already-public
in-tree source, and we intend it for defensive and authorized vulnerability research.
We still split the release in two. Under an open license matching the companion
Windows dataset, we publish the recovered structure (\code{modules},
\code{dispatchers}, \code{ioctl\_codes}, \code{handlers}, \code{gates}, and the
structural views). It records that a module decodes a given command and which gates
sit on the path, but not where an unchecked copy lives. What we hold back is the join
that would point at one, namely the taint sinks, the enrichment verdicts, the
findings, and the ranked unaudited unprivileged surface. In combination they point at
likely unprivileged-reachable primitives across many modules at once, and they stay
dual-use even though each row on its own is re-derivable from source. A column allowlist in the export
tool aborts the build if any withheld field reaches the output. Any live,
unprivileged-reachable, in-threat-model defect we confirm goes through coordinated
disclosure before it appears here, carries no inlined reproducer, and the dataset
release is timed not to front-run those disclosures.

\section{Conclusion}
\label{sec:conclusion}

We set out to make the Linux ioctl surface queryable from source, and the census is
the result. For the \nmodules\ modules an \code{allmodconfig} build compiles across
\nsubtrees\ subtrees, it holds the dispatchers, the \ncodes\ decoded \code{\_IOC}
commands, the \nsinks\ controlled-input sinks, and the \ngates\ permission gates,
recovered per module without ever running the kernel. A second build for \code{arm64}
adds \narmonly\ ioctl module and no new command codes. The threat model the kernel
documents becomes a column, a necessary-condition upper bound rather than a
classifier, which of the \niocmods\ ioctl modules excludes \ncapgated\ behind a hard
capability gate and admits \nreach\ as capability-ungated. What the artifact buys is
persistence: a store that answers questions a bug list cannot, on a schema the
companion Windows census~\cite{ioctlcensuswin,ioctlcensusdata} shares so one query
spans both operating systems. The \ncve-CVE backtest, all in covered subtrees, ties
every recall miss to a single bound, the handler-registration model, and shows the
node-DAC false positives landing exactly where the method predicts. The open work is
concrete and recorded: a
registration model extended to the ops-table dispatch of DRM, V4L2, and ALSA (the
largest remaining handler-miss class), line disciplines, and below-dispatcher callee
helpers; a node-DAC extractor to turn the capability-ungated heuristic into proven
node-openability; and a syzkaller-overlap measurement on the same surface. We
release the structural tier as open data~\cite{linuxioctlcensusdata} and withhold the
targeting tier.

\bibliographystyle{plain}
\bibliography{bib/references}

@misc{syzkaller,
  title        = {syzkaller: an unsupervised coverage-guided kernel fuzzer},
  author       = {{Google}},
  howpublished = {\url{https://github.com/google/syzkaller}},
  note         = {Accessed June 2026. syzbot dashboard:
                  \url{https://syzkaller.appspot.com/}},
  year         = {2026}
}

@misc{ioctlance,
  title  = {{IOCTLance}: Enhanced Vulnerability Hunting in {WDM} Drivers with
            Symbolic Execution and Taint Analysis},
  author = {Lin, Che-Yu},
  year   = {2023},
  note   = {CODE BLUE 2023; presenter handle zeze-zeze.
            \url{https://github.com/zeze-zeze/ioctlance}}
}

@inproceedings{difuze,
  title     = {{DIFUZE}: Interface Aware Fuzzing for Kernel Drivers},
  author    = {Corina, Jake and Machiry, Aravind and Salls, Christopher and
               Shoshitaishvili, Yan and Hao, Shuang and Kruegel, Christopher
               and Vigna, Giovanni},
  booktitle = {Proceedings of the 2017 ACM SIGSAC Conference on Computer and
               Communications Security (CCS)},
  pages     = {2123--2138},
  year      = {2017},
  note      = {\url{https://doi.org/10.1145/3133956.3134069}}
}

@inproceedings{drchecker,
  title     = {{DR. CHECKER}: A Soundy Analysis for {Linux} Kernel Drivers},
  author    = {Machiry, Aravind and Spensky, Chad and Corina, Jake and
               Stephens, Nick and Kruegel, Christopher and Vigna, Giovanni},
  booktitle = {Proceedings of the 26th USENIX Security Symposium},
  pages     = {1007--1024},
  year      = {2017},
  note      = {\url{https://www.usenix.org/conference/usenixsecurity17/technical-sessions/presentation/machiry}}
}

@inproceedings{sdv,
  title     = {Thorough Static Analysis of Device Drivers},
  author    = {Ball, Thomas and Bounimova, Ella and Cook, Byron and Levin,
               Vladimir and Lichtenberg, Jakob and McGarvey, Con and Ondrusek,
               Bohus and Rajamani, Sriram K. and Ustuner, Abdullah},
  booktitle = {Proceedings of the 1st ACM SIGOPS/EuroSys European Conference on
               Computer Systems (EuroSys)},
  pages     = {73--85},
  year      = {2006},
  note      = {Static Driver Verifier (SDV). \url{https://doi.org/10.1145/1217935.1217943}}
}

@misc{screweddrivers,
  title  = {{Screwed Drivers}: Signed, Sealed, Delivered},
  author = {Shkatov, Mickey and Michael, Jesse},
  year   = {2019},
  note   = {Eclypsium research, 2019; presented at DEF CON 27 as ``Get Off the
            Kernel if You Can't Drive''.
            \url{https://eclypsium.com/research/screwed-drivers-signed-sealed-delivered/}}
}

@inproceedings{popkorn,
  title     = {{POPKORN}: Popping {Windows} Kernel Drivers At Scale},
  author    = {Gupta, Rajat and Dresel, Lukas Patrick and Spahn, Noah and
               Vigna, Giovanni and Kruegel, Christopher and Kim, Taesoo},
  booktitle = {Proceedings of the 38th Annual Computer Security Applications
               Conference (ACSAC)},
  pages     = {854--868},
  year      = {2022},
  note      = {\url{https://doi.org/10.1145/3564625.3564631}}
}

@misc{ioctlcensuswin,
  title        = {The {Windows} {IOCTL} Census: A Corpus-Scale, Multi-Architecture
                  Database of the Driver Control-Code Surface},
  author       = {Bommarito II, Michael J.},
  year         = {2026},
  howpublished = {\url{https://arxiv.org/abs/2606.07732}},
  note         = {Companion paper on the shared census schema. arXiv:2606.07732 [cs.SE].}
}

@misc{needles,
  title        = {Needles at Scale: {LLM}-Assisted Target Selection for {Windows}
                  Vulnerability Research},
  author       = {Bommarito II, Michael J.},
  year         = {2026},
  howpublished = {\url{https://arxiv.org/abs/2606.01364}},
  note         = {Companion work on {LLM}-assisted target selection. arXiv:2606.01364 [cs.CR].}
}

@misc{ioctlcensusdata,
  title  = {The Windows {IOCTL} Census (Structural Tier)},
  author = {Bommarito II, Michael J.},
  year   = {2026},
  howpublished = {Hugging Face dataset},
  note   = {\url{https://huggingface.co/datasets/mjbommar/ioctl-census}}
}

@misc{linuxioctlcensusdata,
  title  = {The Linux {IOCTL} Census (Structural Tier)},
  author = {Bommarito II, Michael J.},
  year   = {2026},
  howpublished = {Hugging Face dataset},
  note   = {This work. \url{https://huggingface.co/datasets/mjbommar/linux-ioctl-census}}
}

@misc{linuxthreatmodel,
  title        = {{Linux} kernel security documentation: threat model and
                  security-bugs process},
  author       = {{The Linux Kernel community}},
  howpublished = {\url{https://docs.kernel.org/process/}},
  note         = {Documentation/process/threat-model.rst and
                  security-bugs.rst. As of Linux v7.0-rc7 (the revision this census was built against).},
  year         = {2026}
}

@misc{ioc_uapi,
  title        = {{Linux} uapi: ioctl number encoding},
  author       = {{The Linux Kernel community}},
  howpublished = {\url{https://www.kernel.org/doc/html/latest/userspace-api/ioctl/ioctl-number.html}},
  note         = {The \_IOC direction/type/ordinal/size encoding.},
  year         = {2026}
}

\appendix
\section{Reproducibility and the CVE corpus}
\label{sec:appendix}

\paragraph{Build provenance.} The census is built against Linux \code{v7.0-rc7} with
\code{clang}/\code{libclang} 21.1.8, recorded together with the \code{allmodconfig}
build configuration in the \code{kb\_meta} table. The libclang extractor reads the
\code{compile\_commands.json} that build emits; a \emph{module} is one source file
registering a handler, keyed by \code{module\_sha256 = md5(arch : tree-relative path)}
(Section~\ref{sec:method}). The controlled-input sink allowlist, the depth-six call
closure used for reachability, and the capability-gate predicate
(\code{capable}/\code{ns\_capable}, classified by namespace) are all defined in
Section~\ref{sec:method}; the staging carve-out is the \code{drivers/staging/} path
prefix and debugfs is its registration API. The analysis tables rebuild
byte-identically across two independent runs (Section~\ref{sec:eval}), and the public
export is the structural subset with the targeting-tier tables removed
(Table~\ref{tab:schema}).

\paragraph{CVE corpus selection.} The backtest corpus
(Table~\ref{tab:cve_appendix}) is the set of in-tree Linux ioctl CVEs from
\code{2022} through \code{2025} recorded in the kernel \code{vulns.git} database for
which we could pin a vulnerable handler symbol, a command code, and a source subtree.
We exclude out-of-tree and distro-only issues and bugs whose flaw is not on an ioctl
dispatch path. The corpus is deliberately small and hand-pinned rather than sampled,
so it is a characterization of where the deterministic pass is blind, not a recall
estimate over all kernel CVEs.

\begin{table}[t]
  \centering \footnotesize \setlength{\tabcolsep}{4pt}
  \begin{tabular}{@{}llllccl@{}}
    \toprule
    CVE & subsystem & vulnerable handler & rec. & dec. & adm. & miss reason \\
    \midrule
    CVE-2025-68797 & char & \code{ac\_ioctl} & \checkmark & -- & \checkmark &  \\
    CVE-2022-48872 & misc & \code{fastrpc\_device\_ioctl} & \checkmark & \checkmark & \checkmark &  \\
    CVE-2022-50614 & misc & \code{pci\_endpoint\_test\_ioctl} & \checkmark & \checkmark & \checkmark &  \\
    CVE-2024-41025 & misc & \code{fastrpc\_device\_ioctl} & \checkmark & \checkmark & \checkmark &  \\
    CVE-2025-40117 & misc & \code{pci\_endpoint\_test\_ioctl} & \checkmark & \checkmark & \checkmark &  \\
    CVE-2022-48834 & usb & \code{usbtmc\_ioctl} & \checkmark & \checkmark & \checkmark &  \\
    CVE-2023-53761 & usb & \code{usbtmc\_ioctl} & \checkmark & \checkmark & \checkmark &  \\
    CVE-2024-36917 & block & \code{blk\_ioctl\_discard} & -- & -- & -- & sub-handler \\
    CVE-2024-49994 & block & \code{blk\_ioctl\_discard} & -- & -- & -- & sub-handler \\
    CVE-2022-48899 & gpu drm & \code{virtio\_gpu\_resource\_create\_ioctl} & -- & -- & -- & ops-table \\
    CVE-2024-35786 & gpu drm & \code{nouveau\_gem\_ioctl\_pushbuf} & -- & -- & -- & ops-table \\
    CVE-2024-52559 & gpu drm & \code{msm\_ioctl\_gem\_submit} & -- & -- & -- & ops-table \\
    CVE-2024-53087 & gpu drm & \code{xe\_exec\_ioctl} & -- & -- & -- & ops-table \\
    CVE-2025-21843 & gpu drm & \code{panthor\_ioctl\_dev\_query} & -- & -- & -- & ops-table \\
    CVE-2025-71099 & gpu drm & \code{xe\_oa\_add\_config\_ioctl} & -- & -- & -- & ops-table \\
    CVE-2023-52565 & media & \code{uvc\_query\_v4l2\_menu} & -- & -- & -- & ops-table \\
    CVE-2025-40302 & media & \code{vb2\_ioctl\_remove\_bufs} & -- & -- & -- & ops-table \\
    CVE-2022-48804 & tty & \code{vt\_setactivate} & -- & -- & -- & tty ldisc \\
    CVE-2024-50073 & tty & \code{gsm\_cleanup\_mux} & -- & -- & -- & tty ldisc \\
    CVE-2025-37814 & tty & \code{set\_selection\_kernel} & -- & -- & -- & tty ldisc \\
    CVE-2023-53035 & fs nilfs2 & \code{nilfs\_ioctl\_wrap\_copy} & -- & -- & -- & sub-handler \\
    CVE-2024-35866 & fs smb & \code{cifs\_dump\_full\_key} & -- & -- & -- & sub-handler \\
    \bottomrule
  \end{tabular}
  \caption{The \ncve-CVE backtest corpus (kernel \code{vulns.git}, in-tree ioctl
  CVEs with a known vulnerable handler, command, and subtree, in coverage subtrees).
  rec.\ = handler recovered by the dispatcher pass; dec.\ = command decoded by the
  \code{\_IOC} resolver; adm.\ = admitted by the capability-ungated filter. Miss
  reasons: \emph{ops-table} (DRM/V4L2 dispatch tables), \emph{tty ldisc} (line
  disciplines), \emph{sub-handler} (vulnerable callee below the recovered dispatcher).}
  \label{tab:cve_appendix}
\end{table}

\end{document}